\documentclass[twocolumn,twocolappendix]{aastex631}

\usepackage{amsmath,amssymb}
\usepackage{graphicx}
\usepackage{dcolumn}
\usepackage{bm}
\usepackage{xcolor}

\renewcommand{\bv}[1]{\boldsymbol {#1}}

\newcommand{\rP}{\mathrm{P}}

\newcommand{\rmd}{\mathrm{d}}

\newcommand{\para}{{||}}

\begin{document}

\title{Spacetime Hall-MHD turbulence at sub-ion scales: structures or waves?}

\author{Emanuele Papini}
\email{emanuele.papini@unifi.it}
\affiliation{Dipartimento di Fisica e Astronomia, Universit\`a degli Studi di Firenze, via G. Sansone 1, 50019 Sesto Fiorentino, Italy}
\affiliation{INAF - Osservatorio Astrofisico di Arcetri, Largo E. Fermi 5, 50125, Firenze, Italy}

\author{Antonio Cicone}
\affiliation{Dipartimento di Ingegneria e Scienze dell'Informazione e Matematica, Universit\`a  degli Studi dell'Aquila, via Vetoio 1, 67100 L'Aquila, Italy}
\affiliation{INAF - Istituto di Astrofisica e Planetologia Spaziali, via del Fosso del Cavaliere 100, 00133, Roma, Italy}
\affiliation{Istituto Nazionale di Geofisica e Vulcanologia, Via di Vigna Murata 605, 00143 Roma, Italy}

\author{Luca Franci}
\affiliation{School of Physics and Astronomy, Queen Mary University of London, London E1 4NS, UK}
\affiliation{INAF - Osservatorio Astrofisico di Arcetri, Largo E. Fermi 5, 50125, Firenze, Italy}

\author{Mirko Piersanti}
\affiliation{INAF - Istituto di Astrofisica e Planetologia Spaziali, via del Fosso del Cavaliere 100, 00133, Roma, Italy}

\author{Simone Landi}
\affiliation{Dipartimento di Fisica e Astronomia, Universit\`a  degli Studi di Firenze, via G. Sansone 1, 50019 Sesto Fiorentino, Italy}
\affiliation{INAF - Osservatorio Astrofisico di Arcetri, Largo E. Fermi 5, 50125, Firenze, Italy}

\author{Petr Hellinger}
\affiliation{Astronomical Institute, Czech Academy of Sciences, Bocni II 1401, 141 00 Prague, Czech Republic}

\author{Andrea Verdini}
\affiliation{Dipartimento di Fisica e Astronomia, Universit\`a degli Studi di Firenze, via G. Sansone 1, 50019 Sesto Fiorentino, Italy}
\affiliation{INAF - Osservatorio Astrofisico di Arcetri, Largo E. Fermi 5, 50125, Firenze, Italy}

\received{May 28, 2021}
\accepted{July 7, 2021}
\submitjournal{Astrophysical Journal Letters}

\begin{abstract}
Spatiotemporal properties of {two-dimensional (2D)} Hall-magnetohydrodynamic turbulence at intermediate plasma $\beta=2$ are studied by means of Fast Iterative Filtering, a new technique for the decomposition of nonstationary nonlinear signals. 
Results show that the magnetic energy at sub-ion scales is concentrated in perturbations with frequencies smaller than the ion-cyclotron (IC) frequency and with polarization properties that are incompatible with both kinetic Alfv\'en waves (KAWs) and IC waves.
At higher frequencies, we  clearly identify signatures of both whistler waves and KAWs, however their energetic contribution to the magnetic power spectrum is negligible.
We conclude that the dynamics of {2D} Hall-MHD turbulence at sub-ion scales is mainly driven by localized intermittent structures, with no significant contribution of wavelike fluctuations.

\end{abstract}



\section{\label{sec:level1}INTRODUCTION}

The physical mechanisms underlying the turbulent cascade in the solar wind and other space plasma environments below the proton characteristic scales remain largely unknown.
In particular, it is hotly debated whether the energy cascade at sub-ion scales results from nonlinearly interacting wavelike fluctuations described in terms of low-frequency kinetic Alfv\'en waves (KAW) \citep{2009scheko,2009sahraoui,2012salem,2013chen} and/or whistler waves, eventually complemented by wave-particle interactions \citep[e.g., Landau damping,][]{2010sahraoui,2016sulem}, or from spatially-localized highly-intermittent coherent structures (e.g., current sheets and reconnection sites) where dissipation, heating, and cross-scale energy transfer are enhanced \citep{2012wan,2012perri,2014osman,2017yang,2017cerri,2018camporeale,2020Papini}.
It is difficult to address this question, since the intrinsic spatiotemporal multiscale nature of turbulence makes it difficult  
to identify and isolate the relevant mechanisms in place. 
{At MHD scales and for moderate values of the plasma $\beta$, the turbulent eddies carry a significant part of the energy \citep{2017andres}.
At sub-ion scales, early studies highlighting the spacetime Fourier structure of turbulence hint that the plasma dynamics is dominated by almost static low-frequency features, with no clear signature of waves \citep{2010parashar}.}
Recently, several studies that exploit advanced techniques for the analysis of the spatial properties of turbulence provided evidence that the cross-scale energy transfer at sub-ion scales occurs preferentially in coherent structures through enhanced dissipation \citep{2017yang,2018camporeale,2020Papini}. 
On one hand, such a picture is consistent with the high levels of intermittency measured in the solar wind \citep{2012osman,2013wu}. 
On the other hand, solar wind turbulence at sub-ion scales shown to have properties similar to those of KAWs, such as well defined ratios between the power spectra of magnetic and density fluctuations \citep{2012salem,2013chen} and a reduced magnetic helicity \citep{1982matthaeus} of positive sign at high angles of propagation with respect to the mean magnetic field \citep{2016telloni,2018woodham}.

In this letter, we report results from a high-resolution spatiotemporal multiscale study of fully developed turbulence in a 2D Hall-magnetohydrodynamic (MHD) simulation, {with the aim to assess whether nonlinear wavelike activity at sub-ion scales is energetically relevant to the plasma dynamics.
{Previous studies, comparing Hall-MHD models with hybrid (ion-kinetic and electron-fluid) models \citep{2019papini_turb}  and fully kinetic models \citep{2019gonzalez},
show that Hall-MHD is able to reproduce the main magnetic properties of plasma turbulence at sub-ion scales.}
In turn, hybrid kinetic models well reproduce in-situ observations \citep{2020franci,2021franci}.}
The use of a 2D setup allows for a much higher resolution, compared to a full 3D one, while preserving {many spectral and statistical properties of turbulence \citep{2018franci}. 
It is true, however, that employing a 2D geometry prevents
the onset of 3D intrinsic features of plasma turbulence \citep[e.g., critically balanced turbulence,][]{1995goldreich}. }
The novelty of this work relies in the concurrent use of, (i), a large high-resolution spatiotemporal dataset and, (ii), Fast Iterative Filtering, a novel adaptive technique designed for the decomposition of nonlinear nonstationary signals \citep{2020cicone} (see Appendix \ref{sec:FIF}) which allows to clearly isolate features in the time-frequency domain.

\section{Method}
We performed a simulation of freely-decaying Alfv\'enic turbulence on a box of size $L_x \times L_y = 128 d_i \times 128 d_i$, with $1024^2$ gridpoints and a resolution $\Delta x= \Delta y = d_i/8$ (where $d_i$ is the ion inertial length), by means of a fully-compressible viscous-resistive nonlinear Hall-MHD pseudospectral code \citep[see Appendix \ref{sec:HMHD} and][]{2019papini_turb}. 
The system is initialized with a constant out-of-plane magnetic field $\bv{B}_0 = B_{0} \hat{\bv{z}}$ ($\hat{\bv{z}}$ defines the parallel direction) and  a uniform plasma $\beta = 2$. Moreover, the $xy$-plane is filled with incompressible velocity ($\bv{u}$) and magnetic ($\bv{B}$) fluctuations in energy equipartition and perpendicular to $\bv{B}_0$.
These fluctuations are random-phase sinusoidals of constant amplitude and
with wavenumbers spanning $-8\pi/L_{x(y)}\leq k_{x (y)} \leq 8\pi/L_{x(y)}$. Their initial root-mean-square (rms) amplitude is $u_\mathrm{rms} = B_\mathrm{rms} \simeq 0.24$.

Our analysis begins at the maximum of turbulent activity at $t_0=195\Omega_i^{-1}$ (where $\Omega_i$ is the ion-cyclotron frequency), corresponding to the peak  of the root-mean-square of the current density \citep{2009mininni}. The spacetime analysis is performed in the time interval $[t_0,t_1]=[195\Omega_i^{-1}, 215\Omega_i^{-1}]$, in which all fields are sampled with a cadence of $0.01\Omega_i^{-1}$. 
The resulting fields are defined on a spatiotemporal grid of $1024\times1024\times 2001$ points.
We separate the contribution of fast and slowly evolving features by means of a new approach that uses Fourier Transform (FT) and Fast Iterative Filtering (FIF) \citep{2009lin,2016Cicone} for the spatial and the temporal decomposition, respectively.
We name such mixed approach as the FTFIF decomposition.
The novelty of our analysis is the use of FIF for the temporal decomposition.
FIF is designed to decompose a nonstationary nonlinear signal into a set of Intrinsic Mode Components (IMCs) oscillating around zero but with varying amplitude and frequency, plus a residual or trend. Such decomposition is adaptive, based on the local characteristic frequencies of the signal, and does not make any assumption on the shape of the signal to be extracted. 
As such, FIF offers many advantages over more traditional methods (e.g, FT and Wavelets) that require assumptions on the stationarity and/or linearity of the signal. 
Further details on the FIF technique and on our FTFIF analysis are found in Appendix \ref{sec:FIF} and \ref{sec:FTFIF}.

\section{Results}

\begin{figure*}
    \centering
    \includegraphics[width=\textwidth]{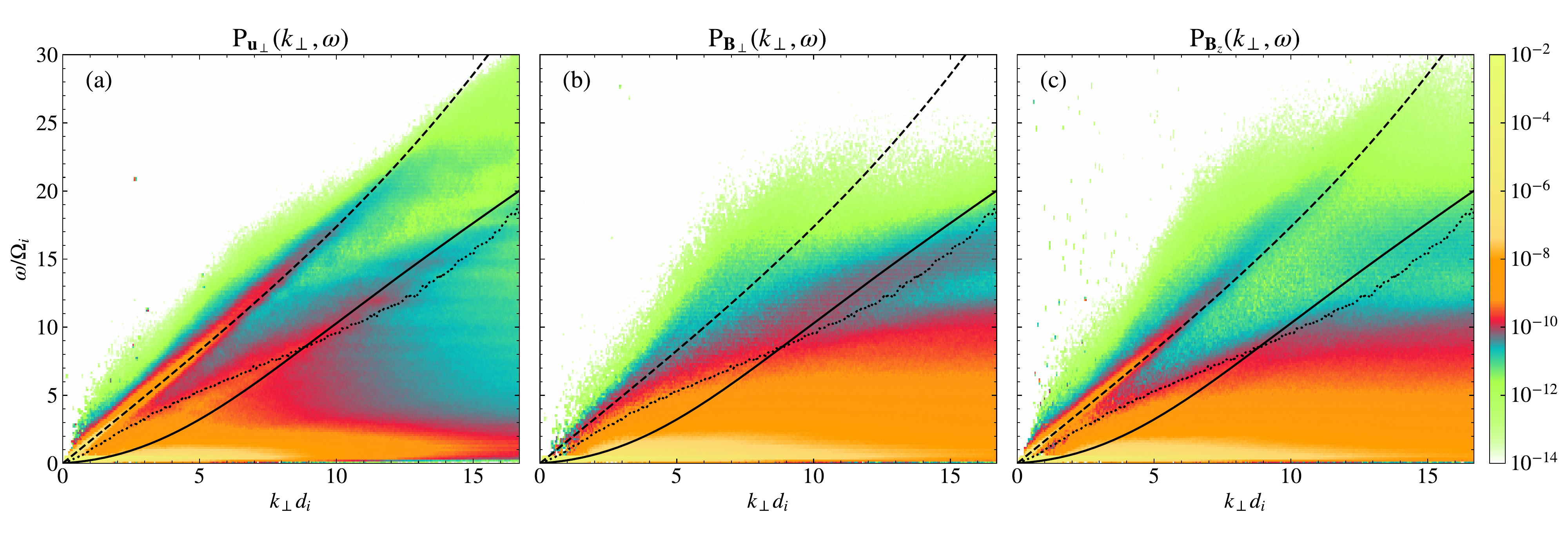}
    \caption{$k\omega$-diagram of the power spectrum obtained from the FTFIF decomposition of the perpendicular component of the fluid velocity (a) and of the perpendicular (b) and parallel  (c) components of the magnetic field at the maximum of turbulent activity. Superimposed is the dispersion relation for FW (dashed) and A/KAW (solid), calculated using the mode ($84.9^\circ$) and the mean ($79.3^\circ$) of the inclination angle $\theta_{Bk} = \arctan(B_z/\sqrt{B_x^2+B_y^2})$,
    respectively.
    The dot-dashed curve denotes the nonlinear-time frequency $\omega_{nl}=2\pi/\tau_{nl} = 2\pi k_\perp \widetilde{u}_e(k_\perp)$\citep{2019papini_turb}.}
    \label{fig:komega_spectrum}
\end{figure*}

Figure \ref{fig:komega_spectrum} shows the $k\omega$-diagram of the power spectrum (or power spectral density) $\mathrm{P}(k_\perp,\omega)$ for both the parallel ($\boldsymbol{B}_z$) and perpendicular ($\boldsymbol{B}_\perp$) magnetic field and for the perpendicular velocity fluctuations ($\boldsymbol{u}_\perp$), as obtained by the FTFIF decomposition. 
{The $k\omega$-diagram of a field is computed by firstly performing a FT in space. Then, a FIF decomposition is performed in time to obtain a set of IMCs for each Fourier mode. Finally, the power and average temporal frequency of each IMC is interpolated to a $k\omega$-grid and summed over all the IMCs (see Appendix \ref{sec:FTFIF}).}
On average, more than 2,000,000 IMCs were extracted per field component.
The $k\omega$-diagram of $\boldsymbol{u}_\perp$ (Fig.\ref{fig:komega_spectrum}a) shows clear signatures of wave activity. Two ridges emanating from the origin nicely follow the theoretical dispersion relations \citep{2016Pucci} for fast/whistler (FW) waves and Alfv\'en/kinetic Alfv\'en waves (A/KAW) respectively.
The dispersion relation for A/KAW is calculated by using the mean ($79.3^\circ$) of the angle $\theta_{Bk}$ between the magnetic field and the $xy$-plane (where the $k$ vectors lie).
Such ridge is broadened because of the excursion of the values of $\theta_{Bk}$ and $\beta$ in the simulation at the maximum of the turbulent activity, which induces changes in the frequencies of the waves throughout the simulation domain.
For FW, the best matching relation is obtained by using the mode ($84.9^\circ$) of $\theta_{Bk}$.
This, together with the fact that the A/KAW ridge shows a wider broadening, suggests that FW are preferentially excited in regions with very small perpendicular magnetic fluctuations where the magnetic field is almost parallel to the mean field $\bv{B}_0$, while KAWs are more widely distributed.
According to its polarizations, the FW ridge is visible also in the $k\omega$-spectrum of $\boldsymbol{B}_z$ and $\rho$ (not shown). The A/KAW ridge is not as visible, since it is partially swamped by the turbulent magnetic activity. 
In the same figure, we report the frequency $\omega_{nl}=2\pi/\tau_{nl}$ corresponding to the characteristic nonlinear turbulent time $\tau_{nl}(k_\perp)=1/(k_\perp \widetilde{u}_e(k_\perp))$, which estimates the rate at which turbulence evolves at a given scale $\ell = 2\pi/k_\perp$ \citep{2019papini_turb} ($\widetilde{u}_e(k_\perp)$ being the amplitude of the electron fluid velocity at that scale). It is reasonable to assume that the area in the $k\omega$-diagram below $\omega_{nl}$ is dominated by the turbulent dynamics whereas, above $\omega > \omega_{nl}$, features (such as the FW ridge and part of the A/KAW ridge) can evolve freely and independently from the main turbulent cascade.

\begin{figure*}
    \centering
    \includegraphics[width=\textwidth]{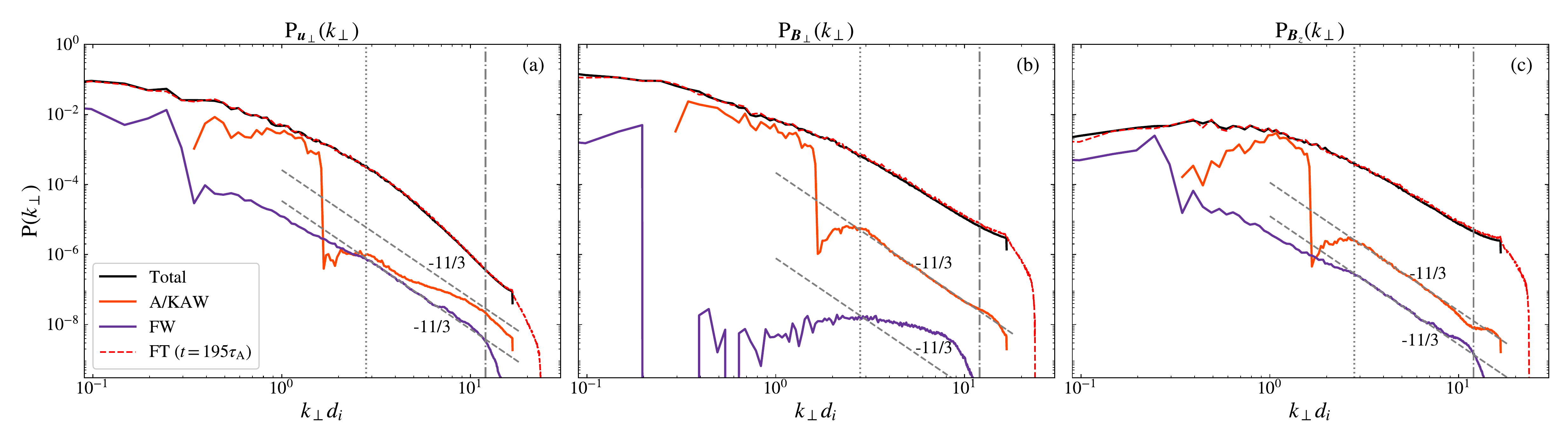}
    \caption{Power spectra $\rP_{\bv{u}_\perp}(k_\perp)$ (a), $\rP_{\bv{B}_\perp}(k_\perp)$ (b), and $\rP_{\bv{B}_z}(k_\perp)$ (c) as obtained by integrating over the FW (violet) and the A/KAW (orange) ridges. 
    The total power spectrum (solid black), obtained by integrating $\rP(k_\perp,\omega)$ over the whole frequency domain, nicely matches the isotropized power spectrum \citep{2015franci_a} obtained from a standard spatial Fourier transform at $t=195\Omega_i^{-1}$ (dashed red). The vertical dotted (dot-dashed) line denotes the position of the break (dissipation) wavenumber.
    }
    \label{fig:branch_spectra}
\end{figure*}

\begin{figure}
    \centering
    \includegraphics[width=\columnwidth,trim={0cm 0.8cm 0cm 0.4cm},clip]{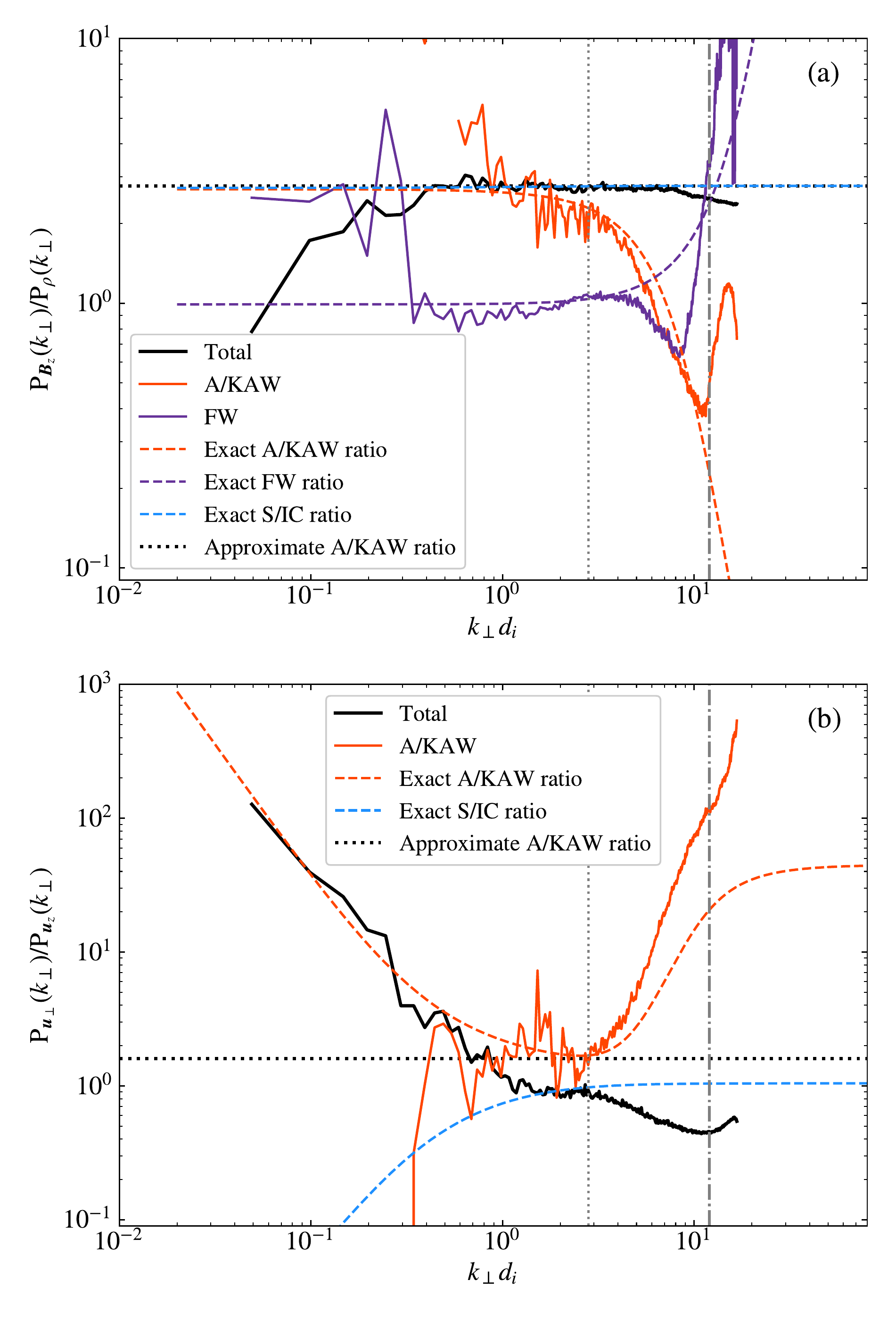}
    \caption{Ratio $\rP_{\bv{B}_z}(k_\perp)/\mathrm{P}_\rho(k_\perp)$ (a) and $\rP_{\bv{u}_\perp}(k_\perp)/\mathrm{P}_{\bv{u}_z}(k_\perp)$ (b) as obtained by integrating the corresponding $k\omega$-spectra over all frequencies (black curve), over the FW ridge (violet), and over the A/KAW ridge (orange). The dashed lines denote the corresponding theoretical ratios as obtained from the wave solutions of the linearized Hall-MHD equations. The horizontal dotted lines denote the approximated KAW ratios (see Appendix \ref{sec:PBKAW}).}
    \label{fig:kaw_ratio_komega}
\end{figure}

To evaluate the energy contribution of FW and A/KAW,
we calculated their $k$-power spectra by integrating in frequency $\mathrm{P}(k_\perp,\omega)$  over the FW and the A/KAW ridge respectively. 
The integration area of the FW ridge encompasses all the frequencies $\omega$ within $10\%$ of the eigenfrequencies of the FW dispersion relation.
Instead, the integration area of the A/KAW ridge was obtained by considering the excursion of the values of $\theta_{Bk}$ and $\beta$ in the simulation domain, which affects the frequencies of KAWs throughout the domain, thus causing the broadening of the ridge.
Results are shown in Fig. \ref{fig:branch_spectra} for $\bv{u}_\perp$, $\bv{B}_\perp$, and $\bv{B}_z$ (but similar results are obtained for the other fields). For all the fields, both the FW (in violet) and KAW (in orange) spectra are energetically negligible at sub-ion scales, being more than one order of magnitude smaller than the total power (in black). 
{The A/KAW spectrum is not negligible at fluid-MHD scales, as expected in the case of Alfv\'enic-turbulence.}
Although energetically irrelevant at sub-ion scales, both the FW and the A/KAW spectra appear to follow well defined power laws.
At around the spectral break ($k_\perp d_i\simeq 2.8$, vertical dotted line) both the FW spectrum (in $\bv{B}_z$) and the A/KAW spectrum (in $\bv{B}_\perp$ and $\bv{B}_z$) sharply transition to a well-defined $-11/3$ slope, which persists until dissipation scales are reached (at $k_\perp d_i\simeq 12$, vertical dot-dashed line).  
The FW spectrum of $\bv{u}_\perp$ also shows a $-11/3$ power-law, because of its polarization properties.
{The existence of a $-11/3$ power-law at sub-ion scales is predicted by few Hall-MHD theoretical models \citep{2004krishan,2007galtier,2012meyrand,2019schekochihin}. Some of their underlying hypothesis, however, are inconsistent with what observed here \citep[e.g. the requirement of $\rP_{\bv{u}}(k_\perp) \gg \rP_{\bv{B}}(k_\perp)$ in][is inconsistent with both the FW and A/KAW spectra]{2004krishan,2007galtier,2019schekochihin}.
This, together with the fact that the whole FW ridge and the A/KAW ridge (although partially) are located at frequencies higher than the nonlinear-time frequency $\omega_{nl}$ (see Fig.\ref{fig:komega_spectrum}), possibly suggests that FW and KAWs may undergo a separate cascade independently from the main cascade.
\begin{figure}
    \centering
    \includegraphics[width=\columnwidth]{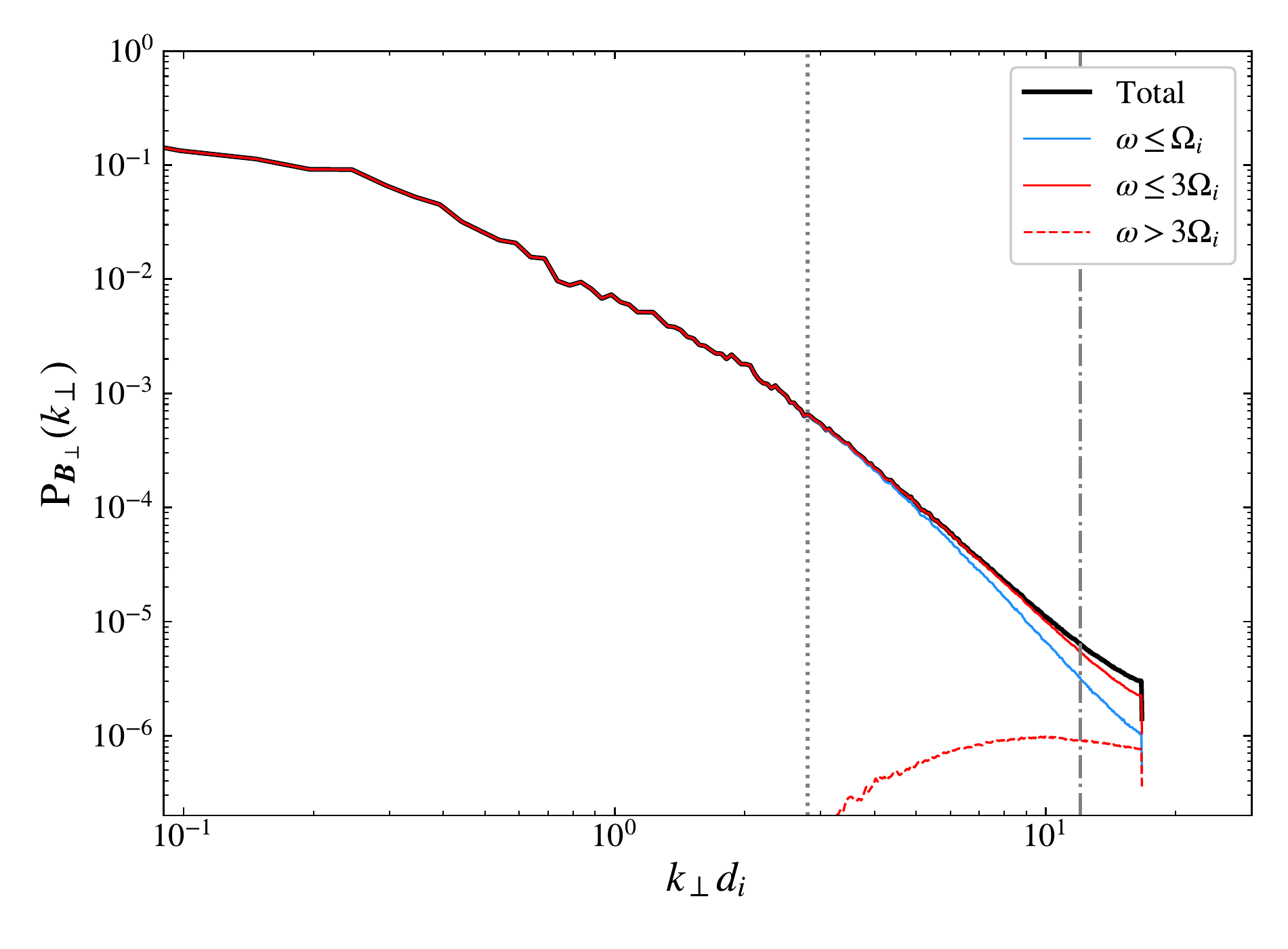}
    \caption{Power spectrum of perpendicular magnetic field fluctuations, obtained by integrating $\rP_{B_\perp}(k_\perp,\omega)$ over all frequencies (black), over low frequencies $\omega\le \Omega_i$ (blue), medium frequencies $\omega \le 3\Omega_i$ (red solid), and high frequencies $\omega > 3\Omega_i$ (red dashed).}
    \label{fig:hor_spectrum}
\end{figure}
We remark that both the FW and the A/KAW spectra remain energetically irrelevant even if the integration area over the corresponding ridge is increased, although the $-11/3$ power-law then tends to disappear.

Instead of being related to FW and/or KAW activity, the energy at sub-ion scales is concentrated at low frequencies. For instance, Fig. \ref{fig:hor_spectrum} shows that the power spectrum of $\bv{B}_\perp$ (black-solid line), mainly results from the energetic contribution of slowly evolving structures and/or perturbations with temporal frequencies smaller than the ion-cyclotron frequency $\Omega_i$ (blue-solid line). Going toward smaller scales ($k_\perp d_i\gtrsim 6$), the contribution of medium frequencies ($\omega\le 3\Omega_i$, red-solid curve) becomes important. The energy contained at higher frequencies (red-dashed curve) is negligible.

\section{DISCUSSION}

The simulation data show that the turbulent {energy} at kinetic scales is {concentrated at low frequencies and it is not clearly} related to KAW activity.
At first glance, these results seem to disagree with previous studies \citep{2013chen_b} which used the ratio between magnetic and density power spectra (the so-called KAW ratios) to detect the presence of KAW activity in the solar wind. 
We deem here that {it may not} be correct to use such ratios as a proxy for KAWs \citep[as also noted by, e.g.,][] {2019Groselj}. {To support such statement}, in Fig. \ref{fig:kaw_ratio_komega}a we report the $\rP_{\bv{B}_z}(k_\perp)/\mathrm{P}_\rho(k_\perp)$ ratio of both the FW (in violet) and the A/KAW (orange) power spectra, together with the ratio of the spectra integrated over all the frequencies (in black). The corresponding theoretical ratios (see Appendix \ref{sec:polarization}), calculated using the exact wave solutions (dashed curves) of the linearized Hall-MHD equations \citep[][]{2016Pucci} nicely match both the FW and the A/KAW ratios from the simulation, further confirming the wave nature of the ridges.
The approximate formula 
$\rP_{\bv{B}_z}(k_\perp)/\mathrm{P}_\rho(k_\perp) = (\Gamma\beta)^2/4$ 
for the KAW ratios (see Appendix \ref{sec:PBKAW}) matches only partially both the exact and the computed KAW ratios, due to the fact that such formula is only valid for large angles $\theta_{Bk} \gtrsim 88^\circ$.
Interestingly, the approximate KAW ratio and the slow/ion-cyclotron (S/IC) exact ratio  match well with the ratio of the total spectra at ion-kinetic scales, thus suggesting the presence of almost-perpendicular low-frequency  KAWs and/or S/IC.
To verify such a possibility, we calculated the ion-velocity polarization ratios (Fig. \ref{fig:kaw_ratio_komega}b). Again, the ratio from the A/KAW ridge nicely match the exact relation at ion-kinetic scales. Moreover, at MHD scales, the ratio of the total spectra (black solid curve) follows the exact A/KAW ratio, as expected since turbulence is Alfv\'enic at those scales.
At ion-kinetic scales, however, neither the approximate KAW (black-dotted line) nor the exact S/IC (blue-dashed curve) ratio match with the computed ratio (solid-black curve). This rules out the presence of (energetically relevant) low-frequency KAWs and S/IC {and point to a different origin for the matching of the approximate KAW ratio with the ratio of the total spectra in Fig. \ref{fig:kaw_ratio_komega}a (e.g. a state of pressure balance at ion-kinetic scales, see Appendix \ref{sec:PBKAW})}}.

{Our results are constrained by the 2D geometry.}
Wave properties are not completely captured in a 2D model, due to the limited access to the parallel wavenumber ($k_{||}$) space. 
{However, we expect that our results are to some extent relevant to a more realistic full 3D case. The} spectral turbulent properties measured in 3D simulations are well reproduced by 2D ones \citep{2018franci}. Moreover, there are regions in our 2D domain where the local mean magnetic field has an in-plane component, thus allowing for the $k_{||}$-space to be partially accessible, especially at small scales. This is confirmed by the existence of A/KAW and FW ridges satisfying dispersion relations with $\theta_{Bk} < 90^\circ$ (see Fig. \ref{fig:komega_spectrum}).
{
Finally and more importantly, theoretical models of KAW turbulence, which require large angles of propagation ($k_{||}\ll k_\perp$), would be captured by our 2D simulation.
In any case, to confirm our findings, we need to extend
this study to a fully 3D numerical setup.}

To summarize, our results show that 2D Hall-MHD turbulence at sub-ion scales is shaped by low frequency features \citep[as reported by][]{2010parashar}. This strongly points to a scenario where the plasma dynamics is dominated by highly-intermittent slowly-evolving structures where enhanced dissipation occurs \citep{2017yang,2018camporeale,2020Papini}.
{Our results indicate that 2D Hall-MHD turbulence on sub-ion
scales is not well described in terms of KAW turbulence phenomenology \citep{2008howes,2009scheko,2013boldyrev}.} 
KAW ratios \citep{2013chen_b} are not a good proxy for discriminating KAW activity, because they only approximate the properties of KAWs and are also sensitive to the presence of S/IC waves and to pressure-balanced structures \citep[see Appendix \ref{sec:PBKAW} and][]{2017Verscharen,2019Groselj}.

Fundamental processes, such as wave-particle resonances and interactions \citep{2006markovskii,2010sahraoui,2016sulem}, as well as nongyrotropic effects \citep{delsarto2016}, may also contribute to dissipation and heating in space plasmas. Such processes are not properly captured by Hall-MHD models. 
In this context, future studies extending the FTFIF analysis to large 3D simulations that retain a fully-kinetic description of plasmas \citep{2015wan,2018franci,2019Groselj,2021franci} will provide fundamental insights into the dynamics and heating in turbulence.

\begin{acknowledgments}
The authors wish to thank the anonymous referee for his constructive review and useful comments.
M. Piersanti thanks the Italian Space Agency for the financial support under the contract ASI ”LIMADOU scienza +” n$^{\circ}$ 2020-31-HH.0.
A.~Cicone is a member of the Italian ``Gruppo Nazionale di Calcolo Scientifico'' (GNCS) of the Istituto Nazionale di Alta Matematica ``Francesco Severi'' (INdAM). 
L.~Franci is supported by the UK Science and Technology Facilities Council (STFC) grant ST/T00018X/1.
This work was supported by the Programme National PNST of CNRS/INSU co-funded by CNES.
We acknowledge partial funding by ``Fondazione Cassa di Risparmio di Firenze'' under the project HIPERCRHEL.
We acknowledge the computing centre of CINECA and INAF, under the coordination of the "Accordo Quadro MoU per lo svolgimento di attività congiunta di ricerca Nuove frontiere in Astrofisica: HPC e Data Exploration di nuova generazione", for the availability of computing resources and support (project INA20\_C6A55)
E.~Papini acknowledges CINECA for awarding access to HPC resources under the ISCRA initiative (grant HP10C2EARF and HP10C4A4M2). 
{The MATLAB code implementing the FIF algorithm is available at \texttt{www.cicone.com}.}
The authors thank L. Matteini and V. Montagud-Camps for useful discussion.
\end{acknowledgments}

\begin{appendix}

\section{Fast Iterarive Filtering}
\label{sec:FIF}
Iterative Filtering (IF) \citep{2009lin,2016Cicone} is a technique for the analysis of nonlinear nonstationary signals.
IF decomposes a given $L^2$ signal $f(t)$ into $N$  simple oscillating functions $\widehat{f}_m(t)$ called intrinsic mode components (IMCs), such that
\begin{equation}
     f(t) = \sum_{m=1}^{N} \widehat{f}_{m}(t) + r(t),
\end{equation}
where $r(t)$ is the residual (or trend) of the decomposition. 
IF uses a low-pass filter, applied iteratively, to extract the moving average of the signal at a given timescale $\tau_m $, in order to isolate a fluctuating component whose average frequency $\omega_m \sim 2\pi/\tau_m$ is well behaved.
Each IMC is given by
\begin{equation}
    \widehat{f}_{m}(t) =\lim_{n\rightarrow\infty} \mathcal{S}_{\tau_m}^n \left [f(t) - \sum_{l=1}^{m-1} \widehat{f}_{l}(t) \right ],
\end{equation}
where $\mathcal{S}_{\tau_{m}} = \mathcal{I} -\mathcal{L}_{\tau_m}$, $\mathcal{I}$ is the identity operator,  $\mathcal{L}_{\tau_m}$ is the integral operator 
associated to the low-pass filter \citep{2020Papini}, and where $n$ denotes the number of times the operator $S_{\tau_{m}}$ is applied.  $\tau_m$ is different for each IMC and increasing with $m$. Therefore, IMCs with increasing $m$ will contain smaller frequencies $\omega_m$.
The low-pass filter operator reads
\begin{equation}
 \mathcal{L}_{\tau_m} [s(t)] = \int_{-\tau_m}^{\tau_m} s(t+\tau)w_m(\tau)\mathrm{d}\tau
\end{equation}
where $w_m(t) \,\epsilon\, [-\tau_m,\tau_m]$ is the kernel function associated to the filter, with compact support $[-\tau_m,\tau_m]$.
Here we employ a Fokker-Plank filter \cite{2016Cicone}. This approach has been recently accelerated in what is known as Fast Iterative Filtering (FIF) \cite{2020cicone}.

\section{Space-time analysis with Iterative Filtering}
\label{sec:FTFIF}
The procedure used to calculate the $k\omega$-diagram is the following.
We consider a field $F(x,y,t)$, with $(x,y)$ defined in a 2D periodic domain on a discrete grid of $N_x\times N_y$ points and with $t\in[t_0,t_1]$.
$F(x,y,t)$ can be either a scalar field (e.g. density) or a vector component (e.g. $B_z$). 
The space-time analysis, employs a mixed approach. We perform a Fourier transform (FT) in space and we then apply FIF in time. The resulting FTFIF decomposition consists of the following steps:
\newcommand{\imf}{\widehat{F}}
\begin{enumerate}
    \item Perform a Fourier transform of $F(x,y,t)$ to obtain its complex Fourier spectrum $\widetilde{F}(k_x^i,k_y^j,t) = \widetilde{F}^{ij}(t)$, with $i (j) = 1,...N_x (N_y)$.
    \item For each real (imaginary) component $F^{ij}(t)$ of $\widetilde{F}^{ij}(t)$, perform an FIF decomposition to obtain $N^{ij}$ IMCs:
    \begin{equation}
     F^{ij}(t) = \sum_{m=1}^{N^{ij}} \widehat{F}_m^{ij}(t)+ r^{ij}(t).
    \end{equation}

    \item For each pair $(i,j)$, increase the orthogonality of the set $\{\widehat{F}_m^{ij}(t)\}$ by calculating the cross-correlation matrix 
    $$ \text{Corr}_{l,n} = \langle \imf_l^{ij} \cdot \imf_n^{ij} \rangle / \sqrt{\langle (\imf_l^{ij})^2\rangle \langle (\imf_n^{ij})^2\rangle} $$
    where 
    $$\langle \imf_l^{ij} \cdot \imf_n^{ij} \rangle = \int_{t_0}^{t_1} \imf_l^{ij}(t) \imf_n^{ij}(t) \mathrm{d}t.$$
    Sum the IMCs for which $\text{Corr}_{l,n} > 0.6$, to obtain a reduced set $\{\widehat{F}_m^{ij}(t)\}$ (with $m=1,M^{ij}$) which is almost orthogonal.
    
    \item For each IMC, calculate its average frequency $\omega_m^{ij}$ and its amplitude 
    $ A_m^{ij} = \sqrt{\langle (\imf_m^{ij})^2\rangle}$.
    At the end you have a set of frequencies and amplitudes
    $ \{\omega_m^{ij},A_m^{ij}\}$, with $m=1,M^{ij}$ and  $i (j)=1,...N_x (N_y) $.
    
    \item Define an (arbitrary) equidistant $k\omega$-grid and interpolate the energy $(A_m^{ij})^2$ of each IMC to the four points nearest to $\left ( k_\perp^{ij} = \sqrt{k_x^{i,2} +k_y^{j,2}}, \omega_m^{ij} \right )$, such that energy is conserved:
    $$
      (A_m^{ij})^2 = \sum_{k_\perp=k_1,k_2}\sum_{\omega = \omega_1,\omega_2} \mathcal{B}\left [(A_m^{ij})^2\right] (k_\perp,\omega),
    $$
    where $\mathcal{B}$ is the interpolation operator and all pair combinations $(k_{1,2},\omega_{1,2})$ denote the coordinates of the four nearest points.
    \item Finally, sum over $i$,$j$, and $m$ to obtain the $k\omega$-power spectrum of $F$
    \begin{equation}
     \mathrm{P}_F(k_\perp,\omega) = \frac{1}{\Delta T}\sum_{i=1}^{N_x} \sum_{j=1}^{N_y}
     \sum_{m=1}^{M^{ij}} \mathcal{B}\left [(A_m^{ij})^2\right] (k_\perp,\omega),
    \end{equation}
    where $\Delta T = t_1-t_0$ is the length of the temporal interval considered.
\end{enumerate}

\section{Hall-MHD model}
\label{sec:HMHD}
The pseudospectral code employed in this work \citep{2019papini_turb} solves the fully-compressible viscous-resistive Hall-MHD equations (in adimensionalized form)
\begin{eqnarray}
 \partial_t \rho   & = & - \nabla\cdot{(\rho \boldsymbol{u})},
  \label{eq:continuity}
  \\
 \rho \rmd_t \boldsymbol{u} 
 & = & -\nabla P + (\nabla\times{\boldsymbol{B}})\times\boldsymbol{B} +
 \nonumber\\
 & & + \mu \left [ \nabla^2\boldsymbol{u} + \nabla(\nabla \cdot {\boldsymbol{u}})/3 \right],
 \label{eq:momentum}\\
  \rmd_t T & = &
  ~(\Gamma  -  1) \! \left \{ - (\nabla\cdot{\boldsymbol{u}})T 
  \! + \!  \eta |\nabla\times\boldsymbol{B}|^2/\rho + \right .\nonumber \\
   & & \left . +  \mu\rho^{-1} \left [ (\nabla\times{\boldsymbol{u}})^2  + 4/3(\nabla\cdot{\boldsymbol{u}})^2\right ] \right \}
  ,   \\
   \partial_t{\boldsymbol{B}} & = & ~\nabla\!\times\!\left ( \boldsymbol{u} \times \boldsymbol{B}\right ) \! + \! \eta \nabla^2 \boldsymbol{B} + \nonumber \\
   &&- d_i/L \nabla\!\times\! [(\nabla\!\times\!\boldsymbol{B})\!\times\!\boldsymbol{B}/\rho ] ,
  \label{eq:induction_hall_adi}
\end{eqnarray}
where $\rmd_t = \partial_t + \bv{u}\cdot\nabla$, $\Gamma=5/3$ is the adiabatic index and $\{\rho,\bv{u},\bv{B},T,P\}$  denote density, ion-fluid velocity, magnetic field, temperature, and pressure respectively. Thermodynamic variables are related through the equation of state $P=\rho T$. 
All quantities are normalized with respect to a characteristic length $L$ (set to be equal to the ion inertial length $d_i$), a plasma density $\rho_0$, a magnetic field amplitude $B_0$, the  Alfv\'en velocity $c_A = B_0/\sqrt{4\pi\rho_0} = \Omega_i d_i$, a pressure $P_0=\rho_0 c_A^2$, and a plasma temperature $T_0 = (k_B/m_i) P_0/\rho_0$. $\Omega_i=e B_0 / m_i c$ is the ion-cyclotron angular frequency and $m_i$ is the ion's mass. With this normalization, the Alfv\'en time is $\tau_A=d_i/c_A = \Omega_i^{-1}$, dynamic viscosity and magnetic resistivity are in units of $d_i c_A \rho_0$ and $d_i c_A$ respectively, and the Hall coefficient $d_i/L=1$ in Eq. (\ref{eq:induction_hall_adi}). 
In the above equations, the magnetic field $\bv{B}$ is expressed in units of the Alfv\'en velocity.
In code units, the electron fluid velocity is $\bv{u}_e = \bv{u}_i - \nabla\times \bv{B}/\rho$.

Equations (\ref{eq:continuity}-\ref{eq:induction_hall_adi}) are solved in a 2.5D periodic cartesian $(x,y)$ domain, by employing Fourier transform for dealiasing and calculating spatial derivatives and by using a 3rd-order Runge-Kutta method for time integration.
In 2.5D codes, the space of the coordinates is 2D, vector quantities retain all three $xyz$-components, and $\partial_z = 0$ for all variables.
The simulation analyzed in the present work employs a 2D box of size $L_x \times L_y = 128 d_i \times 128 d_i$, with $1024^2$ gridpoints and a resolution $\Delta x= \Delta y = d_i/8$. Finally, we set the  plasma $\beta=2P_0/B_0=2$ and $\mu=\eta=0.001$.

\section{Pressure-balanced structures and KAW ratios in Hall-MHD}
\label{sec:PBKAW}

Following \citep{2013boldyrev}, the ratio $(\delta B_{z}(k)/\delta\rho(k))^2$ for KAWs, obtained in the limit $k_z \ll k_\perp$, is given by the formula
\begin{equation}
 \left(\frac{\delta B_{z}(k)/B_0}{\delta\rho(k)/\rho_0}\right)^2 = 
 \left [\frac{v_{T_i}^2}{c_A^2} \left ( 1+\frac{T_e}{T_i}\right) \right ]^2
\end{equation}
where $T_i$ ($T_e$) is the ion (electron) temperature, and $v_{T_i}$ is the thermal speed of the ions.  {The above expression can be adapted to the case of adiabatic HMHD by substituting the ion thermal speed with the adiabatic sound speed ($v_{T_i}\rightarrow c_{s}$) and by using the definition $\beta = (2/\Gamma) c_s^2/c_A^2$} together with $T_e/T_i = \beta_e/\beta_i$ (where we assumed charge neutrality $n_e = n_i=\rho/m_i$) to give
\begin{equation}
 \left(\frac{\delta B_{z}(k)/B_0}{\delta\rho(k)/\rho_0}\right)^2 = \left [\frac{\Gamma}{2}(\beta_i+\beta_e) \right ]^2 = 
 \frac{(\Gamma\beta)^2}{4},
 \label{eq:approx_kaw}
\end{equation}
where $ \beta = \beta_i+\beta_e = (2/\Gamma) c_s^2/c_A^2$ and $c_s^2 = \Gamma P_0/\rho_0$.
{We note that the above formula can be obtained self-consistently from the KAW linear solutions of ideal HMHD (see Eq. \ref{eq:bpar_rat}) in the limit $k_z \ll k_\perp$.}

Eq. \ref{eq:approx_kaw} is also fullfilled by slowly evolving pressure-balanced structures.
In Fourier $k\omega$-space, in the limit of low frequencies and for wavenumbers smaller than the inverse Kolmogorov dissipation scale, Eq. \ref{eq:momentum} reads
\begin{equation}
  \delta P(k) +  \frac{\delta B_{z}(k) B_0}{4\pi} \simeq 0
\end{equation}
to first order in the turbulent fluctuations. By using $\delta P = c_s^2 \delta \rho$, we then obatin
\begin{equation}
 c_s^2 \delta\rho(k) \simeq - \frac{\delta B_{z}(k) B_0}{4\pi}.
\end{equation}
The latter equation may be rearranged to give the ratio between the parallel magnetic fluctuations $\delta B_{z}(k)$ and the density fluctuations $\delta \rho(k)$, to obtain
\begin{equation}
 \frac{\delta B_{z}(k)^2}{\delta \rho(k)^2} = \left (\frac{c_s^2}{c_A^2} \frac{B_0}{\rho_0}  \right )^2 = \left (\frac{\Gamma \beta}{2} \frac{B_0}{\rho_0}  \right )^2
\end{equation}
which, in adimensional units and by using  $\rP_{X}(k)=(\delta X(k))^2/k$, gives the expression
\begin{equation}
 \frac{\rP_{B_z}}{\rP_{\rho}} = \frac{(\Gamma \beta)^2}{4}
\end{equation}
that is equal to Eq. (\ref{eq:approx_kaw}) for the approximated KAW ratios.

\section{Polarization ratios}
\label{sec:polarization}
{In presence of the Hall term and below ion scales ($kd_i\gtrsim 1$), the three MHD wave solutions (fast, slow, and Alfv\'en) are modified. The fast mode turns into a whistler mode (FW), the Alfv\'en mode transitions to a Kinetic Alfv\'en mode (A/KAW), and the slow mode becomes a (compressible) ion-cyclotron mode (S/IC) \citep[e.g.,][]{2000hollweg,2016galtier}.}
The polarization ratios for FW, A/KAW, and S/IC waves are calculated from the eigenmode solutions 
$\{\delta\bv{u}(\omega,\bv{k}),\,\delta\bv{B}(\omega,\bv{k}),\delta\rho(\omega,\bv{k}) \}$ of the linearized Hall-MHD equations. 
The expression for the dispersion relation, $\delta\bv{u}$, and $\delta\bv{B}$ can be found in \citet{2016Pucci}. {Following their convention, here we write the expression for the amplitude of the density
\begin{equation}
 \delta\rho(\omega,\bv{k}) = \left |\rho_0 a \frac{(k_{||}^2c_A^2 - \omega^2)k_\perp \Omega_i}{(\omega^2 - k^2 c_s^2)k_{||}\omega} \right |
\end{equation}
where $\omega$ is the eigenfrequency, $a$ is the perturbation amplitude, }
and $k_{||(\perp)}$ is the parallel (perpendicular) component of the wave vector $\bv{k}$ with respect to the background magnetic field, thus $\tan(\theta_{Bk}) = k_\perp/k_\para$. Note that the definition of parallel direction used here differs from that of the numerical simulation, in which the parallel direction is that of the global mean magnetic field (i.e. the $z$-direction). 

The exact formulas for the polarization ratios shown in this work read
\begin{equation}
 \left (\frac{\delta B_\para \rho_0}{\delta \rho B_0}\right )^2 = 
 \frac{(\omega^2 - k^2 c_s^2)^2}{k^4 c_A^4} ,
 \label{eq:bpar_rat}
\end{equation}
 \begin{equation}
 \left (\frac{\delta B_{\perp}\rho_0}{\delta \rho B_0}\right )^2 = 
 \frac{(\omega^2 - k^2 c_s^2)^2}{(k_\perp/k_{||})^2} 
 \left[ \frac{(\omega/\Omega_i)^2}{(\omega^2 - k_{||}^2c_A^2 )^2} + \frac{1}{k^4 c_A^4}\right],
\label{eq:bperp_rat}
\end{equation}
\begin{equation}
 \left (\frac{\delta u_{\perp}}{\delta u_\para}\right )^2 = 
 \left[ \frac{\omega k_{||}c_A^2 (\omega^2 - k^2 c_s^2)}{\Omega_i k_\perp c_s^2 (\omega^2 - k_{||}^2c_A^2)}\right]^2   
 + \left [\frac{(\omega^2 - k_{||}^2c_s^2)}{k_{||}k_\perp c_s^2}\right]^2,
\label{eq:velocity_rat}
\end{equation}
where the ratio for the desired branch (FW, A/KAW, or S/IC) is obtained by plugging in the corresponding eigenfrequency $\omega$ {\citep[whose analytical expression can be found in][]{1999vocks}}.
The analogous ratios from the numerical simulation have been obtained by using $\delta B_\perp^2/k = \rP_{B_\perp}$, $\delta B_\para^2/k = \rP_{B_z}$, $\delta u_\perp^2/k = \rP_{u_\perp}$, and $\delta u_\para^2/k = \rP_{u_z}$.
Eqs. (\ref{eq:bpar_rat}) and (\ref{eq:velocity_rat}) give the exact ratios used in Fig. \ref{fig:kaw_ratio_komega}. 

\begin{figure}
    \centering
    \includegraphics[width=\columnwidth]{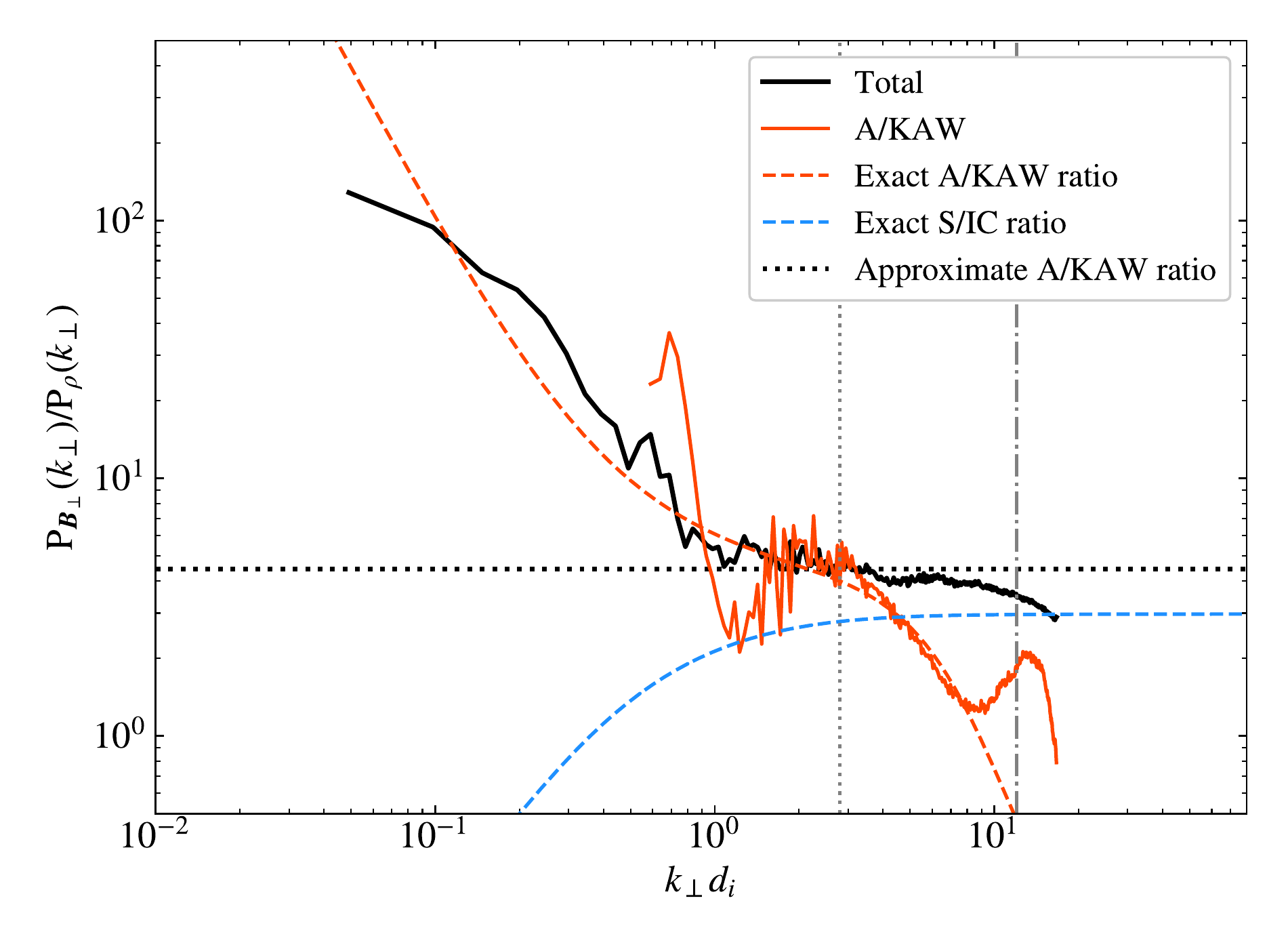}
    \caption{Ratio $\rP_{\bv{B}_\perp}(k_\perp)/\mathrm{P}_{\rho}(k_\perp)$ as obtained by integrating the corresponding $k\omega$-spectra over all frequencies (black curve) and over the A/KAW ridge (orange). The dashed lines denote the corresponding theoretical ratios as obtained from the wave solutions of the linearized Hall-MHD equations. The horizontal dotted lines denote the approximated KAW ratios (see Appendix).}
    \label{fig:kaw_ratio_komega_v}
\end{figure}

In Fig. \ref{fig:kaw_ratio_komega_v} we report the polarization ratios for $\rP_{\bv{B}_\perp}(k_\perp)/\mathrm{P}_{\rho}(k_\perp)$. 
Again, as shown in Fig. \ref{fig:kaw_ratio_komega}, the A/KAW ratio nicely matches the exact relations at ion-kinetic scales. Moreover, at MHD scales, the ratio of the total spectra (black solid curve) follows the exact A/KAW ratio, as expected since turbulence is Alfv\'enic at those scales. At ion-kinetic scales however, this is not the case. 
The approximate KAW ratio (horizontal dotted lines) matches with the ratio of the total spectra for $\rP_{\bv{B}_\perp}/\mathrm{P}_{\rho}$, {(although such match is not as good as in the case of $\rP_{\bv{B}_z}/\mathrm{P}_{\rho}$). However, the presence of energetically relevant low-frequency KAWs has already been ruled out by the $\delta u_\perp^2/\delta u_{z}^2$ ratio. 
This casts some doubts on the use of $\rP_{\bv{B}_\perp}/\mathrm{P}_{\rho}$ as a proxy for KAW activity.}

The approximate relations for the $\rP_{\bv{B}_\perp}/\mathrm{P}_{\rho}$ and the $\rP_{\bv{u}_\perp}/\mathrm{P}_{\bv{u}_z}$ KAW ratios are
\begin{equation}
 \left (\frac{\delta B_{\perp}\rho_0}{\delta \rho B_0}\right )^2 \simeq \frac{\Gamma \beta}{2} 
 \left ( 1 + \frac{\Gamma \beta}{2} \right ),
\end{equation}
\begin{equation}
 \left (\frac{\delta u_{\perp}}{\delta u_\para}\right )^2 \simeq 
\frac{2}{\Gamma\beta} + 1
\end{equation}

\end{appendix}

 \bibliographystyle{aasjournal} 
 \bibliography{bibliography}

\end{document}